# Multimodal Programming in Computer Science with Interactive Assistance Powered by Large Language Model

Rajan Das Gupta[1], Md. Tanzib Hosain[1], M. F. Mridha[1] and Salah Uddin Ahmed[2]

[1] American International University-Bangladesh, Dhaka 1229, Bangladesh
[2] University of South-Eastern Norway, 3199 Borre, Norway
{18-36304-1,20-42737-1}@student.aiub.edu, firoz.mridha@aiub.edu,
salah.ahmed@usn.no

**Abstract.** LLM chatbot interfaces allow students to get instant, interactive assistance with homework, but doing so carelessly may not advance educational objectives. In this study, an interactive homework help system based on DeepSeek R1 is developed and first implemented for students enrolled in a large computer science beginning programming course. In addition to an assist button in a well-known code editor, our assistant also has a feedback option in our command-line automatic evaluator. It wraps student work in a personalized prompt that advances our educational objectives without offering answers straight away. We have discovered that our assistant can recognize students' conceptual difficulties and provide ideas, plans, and template code in pedagogically appropriate ways. However, among other mistakes, it occasionally incorrectly labels the correct student code as incorrect or encourages students to use correct-but-lesson-inappropriate approaches, which can lead to long and frustrating journeys for the students. After discussing many development and deployment issues, we provide our conclusions and future actions.

## 1    Introduction

DeepSeek-V3 and similar Large Language Models (LLMs) have recently been widely available, offering students in introductory Computer Science (CS) courses an alluring substitute for asking for assistance with programming tasks and maybe waiting hours to get it. Nevertheless, although poorly implemented LLMs can help students solve given problems—usually by providing them with accurate answers and explanations—they also let students to avoid the learning that comes with coming up with their own solutions.

To aid students with their assignments in big basic computer science classes, including object-oriented programming (OOP) courses like C, C++, and Java, we developed and implemented an assistance bot at a major private university. We share our experiences here. In the early stages of development, students could only utilize the bot by clicking on a "primary help" button in VS coding, their coding editor. This would result in a pop-up window with instructions on how to complete the homework task given their code that was typed in the editor at that time. We then included a "feedback" option in the course "autoevaluator" command-line tool, allowing all students to utilize the same feedback tool independent of their preferred editor. An au-toevaluator is a tool that automatically assigns grades to students' code contributions according to predetermined standards, such as correctness, efficiency, and style. The IDEs PyCharm, Visual Studio Code, and others do not come with it by default. A unique autoevaluator system and an AI-powered helper bot are both components of the specialized course infrastructure.

After identifying the issue the student is working on, our bot extracts their code and wraps it in a unique Deepseek R1 prompt. We developed this prompt to help Deepseek R1 provide feedback that is comparable to how we typically respond to student questions: assessing whether the student understands the question, identifying which concepts the student might need more explanation of, and assessing whether the student has a plan. If necessary, we then help students by providing conceptual, debugging, or planning support.

We faced several obstacles while developing our system, such as latency constraints and prompt engineering difficulties, which caused our overall system design to shift from a complex, chained analysis-and-response system to a simplified single-prompt method. We provide students with comments on these issues, their design ramifications, and lessons learned from our early deployments, such as the expenses of incorrect bot answers and difficulties evaluating performance.

The core contributions of this paper are:

1. Description of the bot and the reasons behind choosing the current implementation features.
2. Creation of a generalized Deepseek prompt that will work behind the assistance bot's main principle and the development approach of the prompt based on dataset that includes assignments of the course, students complete or incomplete code.

## 2  Related Work

Exciting prospects for improving students' learning experiences are presented by generative models such as Deepseek R1, OpenAI Codex [2], DeepMind Alpha-Code [16], Amazon CodeWhisperer, and GitHub Copilot2. Teachers in a variety of computing education fields have already used these models [10, 8, 12, 6, 23], where they speed up the creation of material and seem to be enhancing the relevant skills that students learn in foundational computer science courses. To improve the comprehensiveness of course materials, researchers have looked into LLMs in areas like automatically creating personalized programming exercises and tutorials [26, 30, 25], generating code explanations [14, 1, 19, 9], providing personalized immediate feedback, and improving programming error messages [15].

Nevertheless, there are several difficulties in including LLMs into introductory computer science classes, particularly those that cover object-oriented programming (OOP) in languages like C, C++, and Java. One major worry is that students may become overly dependent on automation, which has long been a problem in education (like earlier concerns about calculator use [5]), which could hinder their ability to develop critical problem-solving skills. However, new research indicates that this is not always the case for programming help [13]. In addition to additional moral issues related to plagiarism and the proper use of LLM-generated code, the resulting lack of human connection might have negative effects if it is taken too far. A careful and well-rounded approach to integrating LLMs into beginning-level computer science courses is necessary to optimize their advantages while addressing these issues [7, 20, 18].

Like pre-LLM Intelligent Tutoring Systems (review in [4] and example in [27]), students can get immediate, individualized support and guidance by using LLMs as intelligent programming assistants. This allows students to gain a deeper understanding of coding concepts and encourages self-paced learning. Nevertheless, while these generated materials are not always superior, LLMs' capacity to provide customized resources, including tutorials and code samples, not only improves the learning materials that are already accessible but also accommodates students' diverse learning preferences and styles [22].

To improve students' CS education, educators should use LLMs as supplementary tools that strike a balance between automation and human interaction while highlighting the development of important problem-solving skills and proper coding practices. To provide immediate personalized feedback, human-AI pair programming paradigms [28], or tools to support students' development of programming skills [24, 11, 3], such as CodeHelp [17], researchers are increasingly using LLMs as chatbots in courses [9, 29] or online educational websites [21]. This system was created to guide our design and shows how technology-driven methods may revolutionize the way CS fundamentals are taught and learned.

## 3  Methodology

A variety of engagement modes (such as chat chats, Q&A, and one-shot inquiries) and support modes (such as debugging, conceptual scaffolding, and student assessment) were among the methods we investigated to provide LLM capabilities to students. LLMs can help with early programming instruction in a variety of ways, but we decided to focus on one of the more difficult issues we encountered in our university's large introductory course: the scarcity of tutors and other staff members to assist students with their homework. Our objective is to develop a bot that can solve this problem. This well-organized flow demonstrates how an AI-powered system uses an integrated development environment (IDE) to collect data, generate prompts, and respond to requests for code modifications (Fig. 1).

We decided early on to concentrate just on debugging homework assignments to handle this challenge. The possibility of hallucinations (inaccurate or deceptive data presented by the bot), the possibility of students sharing personal information with a third party, and the potential dangers of unmonitored chat discussions are the three main issues that influenced our design decisions. We addressed this by creating a one-click "primary help" interaction mode that restricted participation to a single, brief conversation without follow-up. Although this method reduced the hazards that were identified, it also meant that a crucial teaching tactic—insisting that students explain how they understood the problem—could not be used in this first deployment. We were unable to fully evaluate pupils' issue understanding because of this constraint.

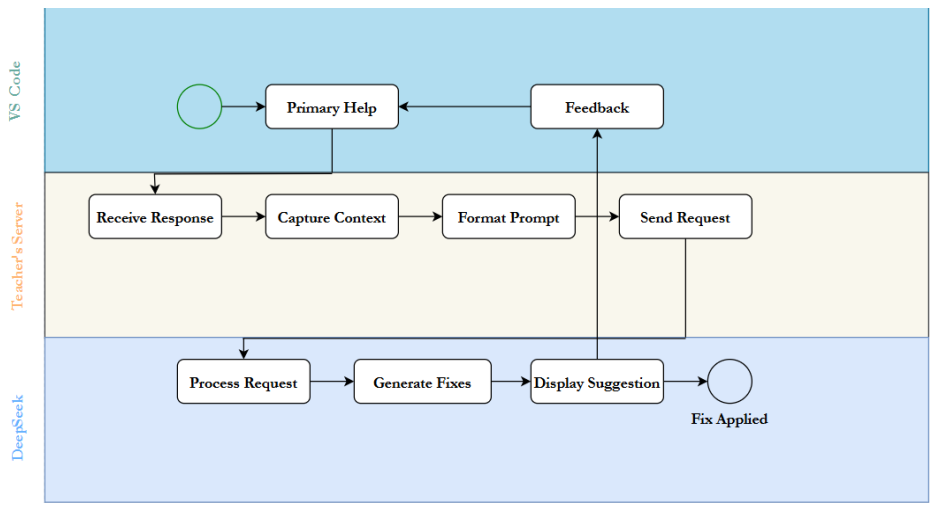

**Fig. 1.** Deepseek R1-powered code fixing pipeline, showcases how an LLM processes programming issues, applies fixes, and generates explanations to improve student learning and debugging efficiency.

To help with code completion and debugging duties, we created an LLM-driven bot that was taught via prompt engineering using DeepSeek R1. Our methodology was based on a collection of student checkpoints—also incomplete code, often including errors—gathered over the course of the previous year, author-generated structures of incomplete code, and a methodically selected set of homework assignments given during the course. We developed our method by employing a custom-built prompt engi-neering system to evaluate our prompt under-development on certain homework problems(Fig.2).

### Exercise: Absolute Value Addition

#### Problem Statement

The Python `operator` module provides functions for basic arithmetic operations. For instance, `operator.add(2, 3)` yields the same result as `2 + 3`, which is 5.

When the `operator` module is imported into the namespace, you can directly use `add(2, 3)` instead of `operator.add(2, 3)`.

Complete the following function to add a to the absolute value of b without using the `abs` function. Only fill in the blanks without altering the rest of the code.

```
def add_abs_value(a, b):
    """Return a + |b| without using abs.
    >>> add_abs_value(2, 3)
    5
    >>> add_abs_value(2, -3)
    5
    >>> add_abs_value(-1, 4)
    3
    >>> add_abs_value(-1, -4)
    3
    """
    if b < 0:
        f = ___
    else:
        f = ___
    return f(a, b)
```

#### Testing Instructions

To test your code, use the following command:

```
python3 ok -q add_abs_value
```

To check for syntax errors without modifying the provided code, use:

```
python3 ok -q add_abs_value.syntax_check
```

#### Solution

The correct implementation is:

```
if b < 0:
    f = sub
else:
    f = add
return f(a, b)
```

Ensure that the conditions are not swapped.

#### Additional Prompts

- Core Concepts
- Code Correction
- Interactive Discussion
- Related Problems

#### Checkpoints

Checkpoints for @.../src (...s)

```
from operator import add, sub
def add_abs_value(a, b):
    """Return a + |b| without using abs.
    >>> add_abs_value(2, 3)
    5
    >>> add_abs_value(2, -3)
    5
    >>> add_abs_value(-1, 4)
    3
    >>> add_abs_value(-1, -4)
    3
    """
    if b < 0:
        f = sub
    else:
        f = add
    return f(a, b)
```

#### Support Prompts

- Chained Help (...s)
- Primary Help

You are assisting a talented student in a programming class at American International University-Bangladesh. The student is working on the provided code and needs guidance.

Consider the following steps to assist effectively:

1. Assess if the student understands the problem.
2. Ensure they grasp the underlying concepts.
3. Help them identify the correct approach.
4. Provide a code template if necessary, ensuring clarity.

The current student's attempt is:

```
if a + is
else:
    f = sub
return f(a, b)
```

Ensure the conditions are correctly applied.

If the student has questions, encourage them to ask. Identify any conceptual misunderstandings and address them without directly altering their code. Use clear explanations and examples.

It appears you understand the problem and concepts, but there's a minor error in your code. In the `add_abs_value` function, f should be assigned a function (either `add` or `sub`), not the result of a function call. Assign f the appropriate function based on the condition, and then call `f(a, b)` in the return statement.

**Fig. 2.** An overview of our unique quick engineering tool. A particular homework question with a handwritten "solution" that includes the appropriate answer and a warning not to switch conditions is on the left; a series of prompts, checkpoints, and outputs are on the right. The "primary help" prompt is triggered in this case, a previous year student's erroneous code is loaded under "checkpoints," and the output is shown beneath the prompt text covered by red rectangle.

Early tests of Deepseek R1 on a limited number of common introductory computer science questions from previous years' courses indicated that Deepseek R1 could offer efficient assistance in a variety of areas, including debugging. It was also significantly more efficient than DeepSeek-V3 and other LLMs, which directed our tuning efforts toward prompt engineering rather than fine-tuning and other techniques. Following a common tutoring pattern, we created a prompt that would attempt to evaluate students' conceptual understanding using the given code and provide ideas for syntactical, logical, or even template-code, but not solutions. A series of actions to take in response to the student code is shown in Fig. 3, which is partially based on our own tutoring procedures.

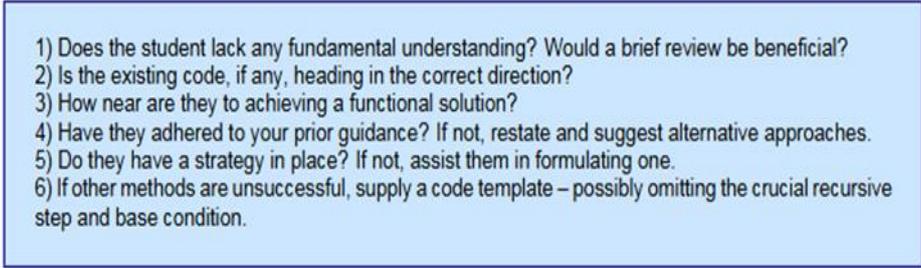

**Fig. 3.** Prompt modeled on personal tutoring processes for evaluating and responding to student code.

We wanted some kind of ongoing interaction even if we didn't let students give the bot direct "chat" messages. To do this, we included in the procedure up to three previous exchanges between the student code and the bot's suggestions (if available). This makes feedback more consistent without requiring face-to-face chat sessions.

The question gives directions such as "Do not offer the student the answer or any code" and "Limit your reaction to a phrase or two at most," in addition to the previously mentioned stages. Fig. 2 displays an early prototype of this question, whereas Fig. 4 represents the full deployed prompt (notice that the %SOLUTION% marker is changed to language relevant to the student's chosen homework assignment).

> You are an advanced R1 tutoring assistant for the introductory computer science course at American International University-Bangladesh. Your role is to guide students in learning programming concepts effectively.
> A student has reached out for assistance. The problem they are working on is described in the next message, followed by the code they have written so far. Please note that the code may include segments related to other questions – focus only on the relevant part. If the student continues to seek help, the conversation will proceed with your further responses and any updates to their code.
> First, evaluate the student's code. If it appears correct and complete, respond with: "Your solution looks good – try running it and share any error messages if they occur!"
> If the code is incomplete or incorrect, consider the following steps:
> 1) Does the student lack any fundamental understanding? Would a brief explanation help?
> 2) Is the existing code moving in the right direction?
> 3) How close is the student to achieving a functional solution?
> 4) Have they followed your previous guidance? If not, rephrase your advice or suggest an alternative approach.
> 5) Do they have a clear plan? If not, assist them in formulating one.
> 6) As a last resort, provide a partial code template, possibly omitting critical components like the base case or recursive logic.
> %SOLUTION%
> Avoid providing direct answers or complete code. If there's an obvious error, point out its location. If there's a conceptual gap, offer a concise explanation. Only suggest recursion if the problem explicitly requires it, and adhere to any functions mentioned in the hints.
> Keep your response brief – one or two sentences at most. Use a Socratic approach to encourage critical thinking, and maintain a friendly and supportive tone.
> Now, let's assist the student!

**Fig. 4.** Fully deployed prompt for comprehensive evaluation and response to student code.

### 3.1 Challenges

We inferred from the research that attempting to initiate conversations using an LLM would be challenging in ways that would be difficult to resolve [31], and we discovered that this was also the case for us. When testing extended dialogue exchanges early on, we found that the likelihood of Deepseek R1 providing a direct answer increased as the discussion went on, confirming our choice to begin with a single-shot request instead of dialogue.

The introduction of the directive "Don't assume a problem needs to use recursion unless it's explicit" helped to eradicate the misconception that loop-based code should be rebuilt using recursion, among other early problems that were resolved with simple changes.

An excessive need to "correct" student code that was previously right, however, is one especially annoying issue that reflects several of the shortcomings we saw early on. One of these homework assignments asked students to use the operators add and sub to calculate a + |b| based on whether b < 0. In an early iteration of our challenge, students were encouraged to consider what to do if b < 0 since Deepseek R1 would consistently query the proper response.

We tried asking Deepseek R1 to first create a complete, correct solution before copying the students' code to solve these and similar problems. We thought that by including these in the prompt response, Deepseek R1 would be less likely to mistakenly identify correct code as having errors.

We were forced to abandon this approach due to the significant additional latency it introduced. In many cases, it added hundreds of tokens to the output, none of which could be streamed to students. This resulted in a time-to-response increase of tens of seconds, which was unacceptable considering our design goal of providing quick interactive feedback.

First, we just included the issue statement in the brief language since we intended to create a prompt that would be useful for a variety of course subjects. But after that, we included a specific note for several assignment problems that required further explanation. For instance, this comment provided a solution to assist the student in completing the operator assignment mentioned above.

### 3.2 Deployment

In two course portions, we implemented a staggered rollout (Fig. 5) using two distinct approaches. As explained in Section 4, the purpose of this first deployment was to comprehend the assistant bot's advantages and disadvantages.

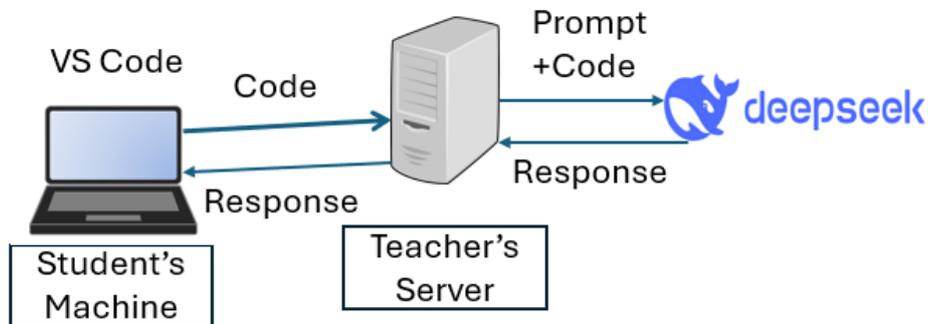

**Fig. 5.** A schematic diagram of the DeepSeek R1-powered bot assistance system deployment.

The assistant bot plugin is activated in the first phase by students clicking a "primary help" button in the VS Code editor toolbar. The plugin collects the student's code, makes an educated guess as to which homework issue the student is working on (since a single file may include numerous problems), and generates a help request.

Students utilize an autoevaluator in the second step, which gathers their code, finds any mistakes, and then creates a request using the data. Both approaches route the queries to a server that is under the control of our teaching team. After enclosing the student's input in a prompt, the server forwards the request to Deepseek R1. The answer from the bot is kept for subsequent examination.

When using our assistance, students are informed that all the code they write will be sent to Deepseek using Microsoft Azure. They are also advised not to put any material (like comments) in their code files that they are uncomfortable sharing.

## 4 Findings

A few days into our academic year, we first made our helper bot VS Code plugin available to 350 students in an experimental section. In week seven, we fully deployed it to roughly 1100 students in both course sections. Instead of being chosen at random, the students in the first experimental phase were mostly selected from a pool of students who had a stronger academic background in math and computers. Students may now obtain automatic feedback while their code is being checked against test cases thanks to our integration of the helper bot with the current autoevaluator tool. This improved accessibility and support for a variety of coding techniques while ensuring continuity for users used to the command-line workflow and minimizing the need to convert to Visual Studio Code.

Students are returning to the bot several times while working on tasks, suggesting that they find it useful. As anticipated, use increases with the approaching assignment due date and is most prevalent in the late afternoons and evenings.

The kinds of errors students experience and the feedback they get seem to have a role in how engaged they are with the helper. Anecdotally, syntax errors and function misuse (when correctly identified by the bot!) are typically resolved with a single request, but conceptual misunderstandings are, predictably, much more difficult. A common pattern for the latter is a student repeatedly asking for help, and the bot responding with slight variations on the same message, until finally, thanks to Deepseek R1's stochasticity, a response contains a crucial new component that allows the student to advance. However, we haven't finished a thorough systematic analysis. (We added the "context" feature mentioned above a few weeks into the semester; it significantly improved answer diversity but did not assist students as much as anticipated, partly because it is not always clear why the previous bot response was insufficient.

Lastly, we saw that the number of support inquiries during office hours had decreased.

### 4.1 Limitations

All things considered, the fact that individual students often utilize our assistance, together with anecdotal feedback and encouraging remarks, shows that students benefit at least somewhat from having the assistant available to them.

We considered Deepseek R1's conceptual explanations to be rather solid on the "successes" side.

The most common mistake patterns on the "failures" side might be categorized as false negatives, where the assistant claims the student code has problems even when it is accurate, and false positives, where the assistant believes the student code is right even though it has issues or provides deceptive answers. In the autoevaluator mode, false negatives are easy to fix; if all test cases pass, the autoevaluator just doesn't ask the helper for input. We want to include similar functionality in our extension.

## 5  Conclusions and Future Directions

Here, we presented first results from an early implementation of an interactive programming aid for introductory computer science courses that is based on Deepseek R1. We discovered several successes, noted some challenges and possible risks, and highlighted solutions and avenues for more comprehensive automated assistance for beginning computer science students.

We want to keep lowering the mistake rate soon by making further adjustments and training. We want to provide the students with the chance to ask for various forms of assistance and rate how beneficial the counsel is to further enhance the setting. While feedback on hint quality might drive a more complex, reinforcement learning-based prompt engineering system, simple answers like "Please elaborate" or "Can you explain it another way?" could give useful context to impact future prompts.

We want to analyze bot use data in the long run to investigate three more areas: effects on student performance, student use knowledge, and student mood assessment. How does the performance of students who use the tool differ from that of students who don't? At what point in the process do students utilize the bot to solve problems? Does the application effectively replace online discussion boards or office hours? How do students feel about interacting with the bot as opposed to a teaching assistant? To evaluate student needs and potential for providing automated help to the next generation of computer scientists, these concerns are essential.

False positives, on the other hand, caused a great deal of trouble for students and may need significant additional work since they were forced to follow recommendations that sometimes took them to places where they found it impossible to relax on their own. One notable instance was an if statement with a misplaced condition, for which the assistant claimed that there was no issue when asked for assistance with the scheme code. The assistant helpfully advised the student to "Ensure all parentheses are matched" when the student astutely replied with the incorrect message obtained from the translator. Consequently, each pair of parentheses had to be manually rebuilt from the beginning by the pupil. Over the course of 15 minutes, a function that had started as a single line was completely redone. The student would have probably gone through a similar process without assistance, which makes this circumstance exceptional but not horrible.

The situations when the recommendations were legitimate but went beyond the assignment's restrictions, such a predetermined template or other limitations, were the most damaging false positives. When given a predetermined template on the one hand and bot recommendations to modify template elements on the other—and therefore conflicting mistakes from the helper and the autoevaluator—students would often alternate between workable answers that met the requirements of one or the other feedback mechanism.

# References

[1] Becker, B.A., Denny, P., Finnie-Ansley, J., Luxton-Reilly, A., Prather, J., Santos, E.A.: Programming is hard–or at least it used to be: Educational opportunities and challenges of AI code generation (2023).

[2] Chen, M., Tworek, J., Jun, H., Yuan, Q., de Oliveira Pinto, H.P., Kaplan, J., Edwards, H., Burda, Y., Joseph, N., Brockman, G., et al.: Evaluating large language models trained on code. arXiv preprint arXiv:2107.03374 (2021).

[3] Cipriano, B.P., Alves, P.: GPT-3 vs Object-Oriented Programming Assignments: An Experience Report. In: Proceedings of the 2023 Conference on Innovation and Technology in Computer Science Education V. 1, pp. 61–67 (2023).

[4] Crow, T., Luxton-Reilly, A., Wuensche, B.: Intelligent Tutoring Systems for Programming Education: A Systematic Review. In: Proceedings of the 20th Australasian Computing Education Conference, pp. 53–62 (2018).

[5] Demana, F., Waits, B.K.: Calculators in Mathematics Teaching and Learning: Past, Present, and Future. In: Learning Mathematics for a New Century, pp. 51–66 (2000).

[6] Denny, P., Becker, B.A., Leinonen, J., Prather, J.: Chat Overflow: Artificially Intelligent Models for Computing Education—Renaissance or Apocalypse? In: Proceedings of the 2023 Conference on Innovation and Technology in Computer Science Education V. 1, pp. 3–4 (2023).

[7] Denny, P., Leinonen, J., Prather, J., Luxton-Reilly, A., Amarouche, T., Becker, B.A., Reeves, B.N.: Promptly: Using Prompt Problems to Teach Learners How to Effectively Utilize AI Code Generators. arXiv preprint arXiv:2307.16364 (2023).

[8] Denny, P., Prather, J., Becker, B.A., Finnie-Ansley, J., Hellas, A., Leinonen, J., Luxton-Reilly, A., Reeves, B.N., Santos, E.A., Sarsa, S.: Computing Education in the Era of Generative AI. arXiv preprint arXiv:2306.02608 (2023).

[9] Donlevy, K.: Harvard to Roll Out AI Professors in Flagship Coding Class for Fall Semester (2023). URL: https://nypost.com/2023/06/30/harvard-to-roll-out-ai-professors-in-flagship-coding-class-for-fall-semester/.

[10] Finnie-Ansley, J., Denny, P., Becker, B.A., Luxton-Reilly, A., Prather, J.: The Robots Are Coming: Exploring the Implications of OpenAI Codex on Introductory Programming. In: Proceedings of the 24th Australasian Computing Education Conference, pp. 10–19 (2022).

[11] Finnie-Ansley, J., Denny, P., Luxton-Reilly, A., Santos, E.A., Prather, J., Becker, B.A.: My AI Wants to Know If This Will Be on the Exam: Testing OpenAI's Codex on CS2 Programming Exercises. In: Proceedings of the 25th Australasian Computing Education Conference, pp. 97–104 (2023).

[12] Hellas, A., Leinonen, J., Sarsa, S., Koutcheme, C., Kujanpää, L., Sorva, J.: Exploring the Responses of Large Language Models to Beginner Programmers' Help Requests. In: Proceedings of the 2023 ACM Conference on International Computing Education Research V. 1 (2023).

[13] Kazemitabaar, M., Chow, J., Ma, C.K.T., Ericson, B.J., Weintrop, D., Grossman, T.: Studying the Effect of AI Code Generators on Supporting Novice Learners in Introductory Programming. In: Proceedings of the 2023 CHI Conference on Human Factors in Computing Systems, pp. 1–23 (2023).

[14] Leinonen, J., Denny, P., MacNeil, S., Sarsa, S., Bernstein, S., Kim, J., Tran, A., Hellas, A.: Comparing Code Explanations Created by Students and Large Language Models. In: Proceedings of the 2023 Conference on Innovation and Technology in Computer Science Education V. 1, pp. 124–130. ACM (2023).

[15] Leinonen, J., Hellas, A., Sarsa, S., Reeves, B., Denny, P., Prather, J., Becker, B.A.: Using Large Language Models to Enhance Programming Error Messages. In: Proceedings of the 54th ACM Technical Symposium on Computer Science Education V. 1. ACM (2023).

[16] Li, Y., Choi, D., Chung, J., Kushman, N., Schrittwieser, J., Leblond, R., Eccles, T., Keeling, J., Gimeno, F., Dal Lago, A., et al.: Competition-Level Code Generation with AlphaCode. Science 378(6624), 1092–1097 (2022).

[17] Liffiton, M., Sheese, B., Savelka, J., Denny, P.: CodeHelp: Using Large Language Models with Guardrails for Scalable Support in Programming Classes (2023).

[18] MacNeil, S., Kim, J., Leinonen, J., Denny, P., Bernstein, S., Becker, B.A., Wermelinger, M., Hellas, A., Tran, A., Sarsa, S., et al.: The Implications of Large Language Models for CS Teachers and Students. In: Proceedings of the 54th ACM Technical Symposium on Computer Science Education, vol. 2 (2023).

[19] MacNeil, S., Tran, A., Hellas, A., Kim, J., Sarsa, S., Denny, P., Bernstein, S., Leinonen, J.: Experiences from Using Code Explanations Generated by Large Language Models in a Web Software Development E-Book. In: Proceedings of the 54th ACM Technical Symposium on Computer Science Education V. 1, pp. 931–937 (2023).

[20] Mirhosseini, S., Henley, A.Z., Parnin, C.: What is Your Biggest Pain Point? An Investigation of CS Instructor Obstacles, Workarounds, and Desires. In: Proceedings of the 54th ACM Technical Symposium on Computer Science Education V. 1, pp. 291–297 (2023).

[21] Ofgang, E.: What is Khanmigo? The GPT-4 Learning Tool Explained by Sal Khan. Tech & Learn (2023).

[22] Pardos, Z.A., Bhandari, S.: Learning Gain Differences Between ChatGPT and Human Tutor Generated Algebra Hints. arXiv preprint arXiv:2302.06871 (2023).

[23] Prather, J., Denny, P., Leinonen, J., Becker, B.A., et al.: Transformed by Transformers: Navigating the AI Coding Revolution for Computing Education. In: Proceedings of the 2023 Conference on Innovation and Technology in Computer Science Education V. 2, pp. 561–562 (2023).

[24] Prather, J., Reeves, B.N., Denny, P., Becker, B.A., et al.: "It's Weird That It Knows What I Want": Usability and Interactions with Copilot for Novice Programmers. arXiv preprint arXiv:2304.02491 (2023).

[25] Reeves, B., Sarsa, S., Prather, J., Denny, P., Becker, B.A., et al.: Evaluating the Performance of Code Generation Models for Solving Parsons Problems. In: Proceedings of the 2023 Conference on Innovation and Technology in Computer Science Education V. 1, pp. 299–305 (2023).

[26] Sarsa, S., Denny, P., Hellas, A., Leinonen, J.: Automatic Generation of Programming Exercises and Code Explanations Using Large Language Models. In: Proceedings of the 2022 ACM Conference on International Computing Education Research-Volume 1, pp. 27–43 (2022).

[27] Suzuki, R., Soares, G., Head, A., Glassman, E., Reis, R., Mongiovi, M., D'Antoni, L., Hartmann, B.: TraceDiff: Debugging Unexpected Code Behavior Using Trace Divergences. In: 2017 IEEE Symposium on Visual Languages and Human-Centric Computing (VL/HCC), pp. 107–115. IEEE (2017).

[28] Wu, T., Koedinger, K., et al.: Is AI the Better Programming Partner? Human-Human Pair Programming vs. Human-AI Pair Programming. arXiv preprint arXiv:2306.05153 (2023).

[29] Wu, Y.C., Petersen, A., Zhang, L.: Student Reactions to Bots on Course Q&A Platforms. In: Proceedings of the 27th ACM Conference on Innovation and Technology in Computer Science Education Vol. 2, p. 621 (2022).

[30] Yuan, Z., Liu, J., Zi, Q., Liu, M., Peng, X., Lou, Y.: Evaluating Instruction-Tuned Large Language Models on Code Comprehension and Generation. arXiv preprint arXiv:2308.01240 (2023).

[31] Zamfirescu-Pereira, J.D., Wei, H., Xiao, A., Gu, K., Jung, G., Lee, M.G., Hartmann, B., Yang, Q.: Herding AI Cats: Lessons from Designing a Chatbot by Prompting GPT-3. In: Proceedings of the 2023 ACM Designing Interactive Systems Conference, pp. 2206–2220 (2023). URL: https://doi.org/10.1145/3563657.3596138.